\documentclass[leqno,fleqn,11pt]{article}

\usepackage[latin1]{inputenc}   
\usepackage{xspace}

\usepackage{graphicx}           
\usepackage{subfigure}
\usepackage{latexsym}           
\usepackage{amsmath}
\usepackage{amssymb}
\usepackage{amsfonts}
\usepackage{fullpage}
\usepackage{charter}

\usepackage{theorem}            

\usepackage[sort&compress,numbers]{natbib}

\newtheorem{thm}{Theorem}[section]
\newtheorem{cor}[thm]{Corollary}

\newtheorem{prop}[thm]{Proposition}

\newtheorem{defn}[thm]{Definition}

\newenvironment{proof}{{\bf Proof:} \rm}{\hfill $\square$ \medskip\\}

\newcommand{\old}[1]{{}}

\newcommand{\UDG}{{\rm{UDG}}}
\newcommand{\paz}{G_\lambda^\theta}

\begin{document}

\title{On a family of strong geometric spanners that admit local routing
strategies\thanks{Research supported in part by NSERC, MITACS, MRI,
and HPCVL.}}
\author{Prosenjit Bose, Paz Carmi, Mathieu Couture, Michiel Smid, Daming Xu\\School of Computer Science, Carleton University}

\maketitle

\begin{abstract} We introduce a family of directed geometric graphs, denoted $\paz$,
that depend on two parameters $\lambda$ and $\theta$.  For $0\leq
\theta<\frac{\pi}{2}$ and $\frac{1}{2} < \lambda < 1$, the $\paz$
graph is a strong $t$-spanner, with
$t=\frac{1}{(1-\lambda)\cos\theta}$. The out-degree of a node in the
$\paz$ graph is at most $\lfloor2\pi/\min(\theta,
\arccos\frac{1}{2\lambda})\rfloor$. Moreover, we show that routing
can be achieved locally on $\paz$. Next, we show that all strong
$t$-spanners are also $t$-spanners of the unit disk graph.
Simulations for various values of the parameters $\lambda$ and
$\theta$ indicate that for random point sets, the spanning ratio of
$\paz$ is better than the proven theoretical bounds.
\end{abstract}




\section{Introduction}

A graph $G$ whose vertices are points in the plane and edges are
segments weighted by their length is a {\em geometric graph}. A
geometric graph $G$ is a $t$-spanner (for $t\geq 1$) when
the weight of the shortest path in $G$ between any pair of points
$a,b$ does not exceed $t\cdot |ab|$ where $|ab|$ is the Euclidean
distance between $a$ and $b$. Any path from $a$ to $b$ in $G$ whose
length does not exceed $t\cdot |ab|$ is a $t$-spanning path. The
smallest constant $t$ having this property is the {\em spanning
ratio} or {\em stretch factor} of the graph. A $t$-spanning path
from $a$ to $b$ is {\em strong} if the length of every edge in the
path is at most $|ab|$. The graph $G$ is a {\em strong} $t$-spanner
if there is a strong $t$-spanning path between every pair of
vertices.

The spanning properties of various geometric graphs have been
studied extensively in the literature (see the book by
\citet{smid07} for a comprehensive survey on the topic). We are
particularly interested in spanners that are defined by some
proximity measure or emptiness criterion (see for example
\citet{bose06a}). Our work was initiated by \citet{kranakis04} who
introduced a new geometric graph called Half-Space Proximal (HSP).
Given a set of points in the plane, HSP is defined as follows. There
is an edge oriented from a point $p$ to a point $q$ provided there
is no point $r$ in the set that satisfies the following conditions:

\begin{enumerate}
\setlength{\itemsep}{-1mm}
\item $|pr|<|pq|$,
\item there is an edge from $p$ to $r$ and
\item $q$ is closer to $r$ than to $p$.
\end{enumerate}

The authors show that this graph has maximum
out-degree\footnote{Theorem 1 in \cite{kranakis04}} at most~6. The
authors also claim that HSP has an upper bound of $2\pi +1$ on its a
stretch factor\footnote{Theorem 2 in \cite{kranakis04}} and that
this bound is tight\footnote{Construction in Figure 2 in
\cite{kranakis04}}. Unfortunately, in both cases, we found
statements made in their proofs of both the upper and lower bounds
to be erroneous or incomplete as we outline in
Section~\ref{section-comparison}. However, in reviewing their
experimental results as well as running some of our own, although
their proofs are incomplete, we felt that the claimed results might
be correct. Our attempts at finding a correct proof to their claims
was the starting point of this work. Although we have been unable to
find a correct proof of their claims, we introduce a family of
directed geometric graphs that approach HSP asymptotically and
possess several other interesting characteristics outlined below.

In this paper, we define a family $\paz$ of graphs. These are
directed geometric graphs that depend on two parameters $\lambda$
and $\theta$. We show that each graph in this family has bounded
out-degree and is a strong $t$-spanner, where both the out-degree
and $t$ depend on $\lambda$ and $\theta$. Furthermore, graphs in
this family admit local routing algorithms that find strong
$t$-spanning paths. Finally, we show that all strong $t$-spanners
are also spanners of the unit-disk graph, which are often used to
model adhoc wireless networks.

The remainder of this paper is organized as follows. In
Section~\ref{section-paz}, we introduce the $\paz$ graph and prove
its main theoretical properties. In Section~\ref{section-comparison}, we
compare the $\paz$ graph to HSP. In Section~\ref{section-udg}, we show
that by intersecting the $\paz$ graph with the unit disk graph, we
obtain a spanner of the unit disk graph. In
Section~\ref{section-simres}, we present some simulation results about
the $\paz$ graph.

\section{The family $\paz$ of graphs}\label{section-paz}

In this section, we define the $\paz$ graph and prove that it is a
strong $t$-spanner of bounded out-degree. We first introduce some
notation. Let $P$ be a set of points in the plane, $0\leq
\theta<\frac{\pi}{2}$ and $\frac{1}{2} < \lambda < 1$.

\vspace{-2mm}
\begin{defn} The \emph{$\theta$-cone$(p,r)$} is the cone of
angle $2\theta$ with apex $p$ and having the line through $p$ and
$r$ as bisector.
\end{defn}

\vspace{-5mm}
\begin{defn}
The \emph{$\lambda$-half-plane$(p,r)$} is the half-plane containing
$r$ and having as boundary the line perpendicular to $\overline{pr}$
and intersecting $\overline{pr}$ at distance
$\frac{1}{2\lambda}|pr|$ from $p$.
\end{defn}

\begin{figure}
\begin{center}
\includegraphics{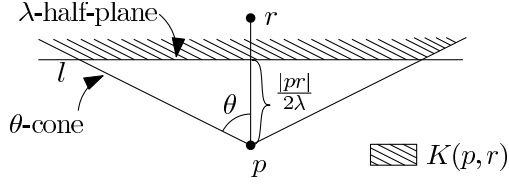}
\end{center}
\caption{The destruction region of $r$ with respect to
$p$.}\label{fig-destruction}
\end{figure}

\vspace{-5mm}
\begin{defn}
The \emph{destruction region} of $r$ with respect to $p$, denoted
$K(p,r)$, is the intersection of the $\theta$-cone$(p,r)$ and the
$\lambda$-half-plane$(p,r)$ (see Figure~\ref{fig-destruction}).
\end{defn}

The directed graph $\paz(P)$ is obtained by the following algorithm.
For every point $p \in P$, do the following:
\begin{enumerate}
\setlength{\itemsep}{-1mm}
\item Let $N(p)$ be the set $P \setminus \{p\}$.
\item
\label{jumpLable} Let $r$ be the point in $N(p)$ which is closest to $p$.
\item Add the directed edge $(p,r)$ to $\paz(P)$.
\item
\label{removeLabel}
Remove all $q \in K(p,r)$ from $N(p)$
(i.e., $N(p) \leftarrow N(p) \setminus K(p,r) $).
\item If $N(p)$ is not empty go to \ref{jumpLable}.
\end{enumerate}
The graph computed by this algorithm can alternatively be defined in the
following way:

\begin{defn}
The directed graph $\paz(P)$ is the graph having $P$ as vertex set and
there is an edge $(p,q) \in \paz(P)$ iff there is no point $r\in P$,
such that $|pr| \leq |pq|$, $(p,r)$ is an edge of $\paz(P)$ and $q
\in K(p,r)$, (ties on the distances are broken arbitrarily). Such a
point $r$ is said to be a \emph{destroyer} of the edge $(p,q)$.
\end{defn}

\subsection{Location of Destroyers}

What prevents the directed pair $(p,q)$ from being an edge in $\paz$? It
is the existence of one point acting as a destroyer. Given two
points $p,q$, where can a point lie such that it acts as the
destroyer of the edge $(p,q)$? In this subsection, we describe the
region containing the points $r$ such that $q\in K(p,r)$. This
region is denoted $\overline{K}(p,q)$. In other words,
$\overline{K}(p,q)$ is the description of all the locations of
possible destroyers of an edge $(p,q)$.

\begin{figure}[h]
\begin{center}
\includegraphics{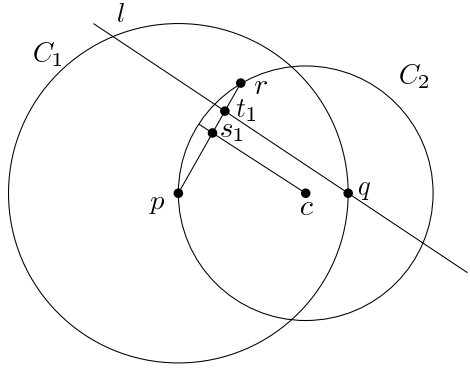}
\end{center}\caption{The location of a point $r$ destroying the edge
$(p,q)$.}\label{fig-destruction1}
\end{figure}
\begin{prop}\label{prop-lune}
Let $R(p,q,\lambda)$ be the intersection of the disks $C_1$ centered
at $p$ with radius $|pq|$ and $C_2$ centered at $c=p + \lambda(q-p)$
with radius $\lambda|pq|$. If $q \in K(p,r)$ and $|pr| \leq |pq|$,
then $r\in R(p,q,\lambda)$.
\end{prop}
\begin{proof} If $r$ destroyed $(p,q)$, then $|pr| \leq |pq|$. Therefore,
$r$ is in $C_1$. To complete the proof, we need to show that $r$ is
in $C_2$. We begin by considering the case when $q$ lies on the line
$l$ which is the boundary of $\lambda$-half-plane$(p,r)$ (see
Figure~\ref{fig-destruction1}). Let $s_1$ be the midpoint of
$\overline{pr}$, $t_1$ the intersection of $l$ with $\overline{pr}$
and $c'$ the intersection of $\overline{pq}$ with the bisector of
$\overline{pr}$. Since the triangles $\triangle pt_1q$ and
$\triangle ps_1c'$ are similar, this implies that
$$
|pc'|  =  |pq|\frac{|ps_1|}{|pt_1|}  =  |pq|\frac{|pr|}{2|pt_1|}  =
 |pq|\frac{2\lambda|pr|}{2|pr|}  =  \lambda|pq|  =  |pc|.
$$
Therefore, $c'=c$, which implies that $|cr|=|cp|$ thereby proving
that $r$ is on the boundary of $C_2$.

In the case when $q$ is not on $l$, then we have $|pc'|<|pc|$ and
$r$ lies on a circle centered at $c'$ going through $p$. Therefore,
$r$ is contained in $C_2$, which completes the proof.\end{proof}

The following proposition follows directly from the definition of
$K(p,r)$.

\begin{prop}\label{prop-cone}If $q \in K(p,r)$, then $\angle qpr\leq\theta$.
\end{prop}

Combining Proposition~\ref{prop-lune} and
Proposition~\ref{prop-cone}, we get:

\begin{prop}\label{prop-destroyer-location}Let $\overline{K}(p,q)$ be the intersection of
$R(p,q,\lambda)$ with the $\theta$-cone$(p,q)$. If $q\in K(p,r)$ and
$|pr| \leq |pq|$, then $r\in \overline{K}(p,q)$.
\end{prop}

\subsection{The Stretch Factor of $\paz$}

\begin{figure}
\centering
\includegraphics[scale=0.8]{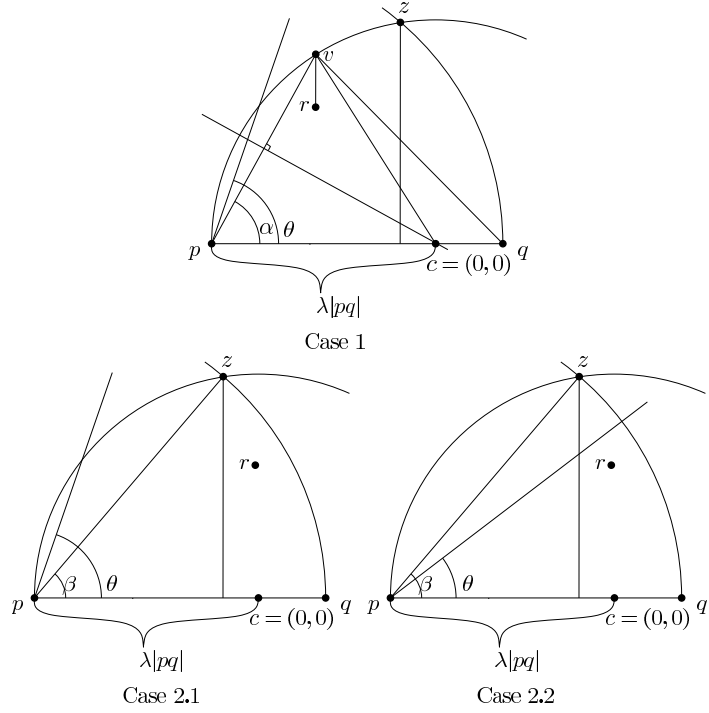}
\caption{Cases for the proof of
Theorem~\ref{thm-stretch}.}\label{thmcases}
\end{figure}

\begin{thm}\label{thm-stretch}For $0\leq
\theta<\frac{\pi}{2}$ and $\frac{1}{2} < \lambda < 1$, the $\paz$
graph is a strong $t$-spanner, with
$t=\frac{1}{(1-\lambda)\cos\theta}$.
\end{thm}
\begin{proof} Let $P$ be a set of points in the plane, $p,q\in P$ and
$d_G(p,q)$ be the length of the shortest path from $p$ to $q$ in
$\paz(P)$. We show by induction on the rank of the distance $|pq|$
that $d_G(p,q)\leq t|pq|$.

\ \\ \textbf{Base case:} If $p$ and $q$ form a closest pair, then the edge $(p,q)$ is in $\paz(P)$
by definition. Therefore, $d_G(p,q)=|pq|\leq t|pq|$.\\
\ \\
\textbf{Inductive case:} If the edge $(p,q)$ is in $\paz(P)$, then
$d_G(p,q)=|pq|\leq t|pq|$ as required. We now address the case when
$(p,q)$ is {\em not} in $\paz(P)$. By
Proposition~\ref{prop-destroyer-location}, there must be a point
$r\in \overline{K}(p,q)$ with $|pr| < |pq|$ that is destroying $(p,q)$
and such that the edge $(p,r)$ is in $\paz(P)$. Since $r\in \overline{K}(p,q)$ and
$|pr|<|pq|$, we have that $|rq| < |pq|$. By the inductive
hypothesis, we have $d_G(r,q)\leq t|rq|$.

Let $z$ be the intersection of the boundaries of the disks $C_1$ and
$C_2$ defined in Proposition~\ref{prop-lune}. We assume, without
loss of generality, that $c$ is the origin and that points $p,q$ are
on the $x$-axis with $p$ to the left of $q$ as depicted in Figure
\ref{thmcases}. The remainder of the proof addresses two cases,
depending on whether or not $r_x\leq z_x$ (the notation $p_x$
denotes the $x$-coordinate of a point $p$).\\
\ \\
\textbf{Case 1: $r_x  \leq z_x$.} Let $v\in \overline{K}(p,q)$ be
the point with the same $x$-coordinate as $r$ and having the
greatest $y$-coordinate. In other words, $v$ is the highest point in
$\overline{K}(p,q)$ that is strictly above $r$. We have:
\begin{eqnarray*}
d_G(p,q) & \leq & |pr| + d_G(r,q)\\
       & \leq & |pr| + t|rq|\mbox{\rm {(ind. hyp.)}}\\
       & \leq & |pv| + t|vq|.
\end{eqnarray*}
Now, let $\alpha=\angle vpq\leq\theta$ We express $|pv|$ and $|vq|$
as a function of $\cos\alpha$. Consider the triangle
$\triangle(pvc)$ and note that $|vc|=|pc|$ by construction. We have
\[ |pv|= 2\lambda|pq|\cos\alpha
\]
and, from the law of cosines,
\begin{eqnarray*}
|vq|^2 &=& |pv|^2 + |pq|^2 - 2|pv||pq|\cos\alpha\\
& = & 4\lambda^2|pq|^2\cos^2\alpha + |pq|^2 - 4\lambda|pq|^2\cos^2\alpha\\
& = & |pq|^2(4\lambda^2\cos^2\alpha - 4\lambda\cos^2\alpha + 1)
\end{eqnarray*}
which implies that:
\begin{eqnarray*}
d_G(p,q) & \leq & 2\lambda|pq|\cos\alpha + t|pq|\sqrt{4\lambda^2\cos^2\alpha - 4\lambda\cos^2\alpha + 1}\\
& = & |pq|(2\lambda\cos\alpha + t\sqrt{4\lambda^2\cos^2\alpha -
4\lambda\cos^2\alpha + 1}).
\end{eqnarray*}
Therefore, we have to show that:
\[ t \geq  2\lambda\cos\alpha + t\sqrt{4\lambda^2\cos^2\alpha -
4\lambda\cos^2\alpha + 1} ,
\]
which can be rewritten as
\[ t \geq \frac{2\lambda\cos\alpha}{1-\sqrt{4\lambda^2\cos^2\alpha
- 4\lambda\cos^2\alpha + 1}} .
\]
Since $\alpha\leq\theta <\pi/2$ implies
$\cos\theta\leq\cos\alpha$, by straightforward algebraic
manipulation we have that
\begin{eqnarray*}
\frac{1}{(1-\lambda)\cos\alpha} & \geq &
\frac{2\lambda\cos\alpha}{1-\sqrt{4\lambda^2\cos^2\alpha -
4\lambda\cos^2\alpha + 1}} .
\end{eqnarray*}


\ \\
\textbf{Case 2: $r_x  > z_x$.} Let $\beta=\angle zpq$. We first
compute the value of $\cos\beta$. From the definition of $C_1$ and
$C_2$, we have
$$z_x^2+z_y^2=\lambda^2|pq|^2$$
and
$$(z_x-p_x)^2+z_y^2=|pq|^2.$$
Therefore, since $p_x=-\lambda|pq|$, we have $z_x= \frac{|pq|(1 -
2\lambda^2)}{2\lambda}$ which implies
$$\cos\beta=\frac{\lambda|pq|+z_x}{|pq|}=\lambda+\frac{1-2\lambda^2}{2\lambda}=\frac{1}{2\lambda}.$$
We need to consider two subcases, depending on whether or not $\beta\leq\theta$.\\
\ \\ \textbf{Case 2.1: $\beta\leq\theta$.} In this case, we have:
\begin{eqnarray*}
d_G(p,q) & \leq & |pr| + d_G(r,q)\\
       & \leq & |pr| + t|rq|\mbox{\rm {(ind. hyp.)}}\\
       & \leq & |pz| + t|zq|\\
       & = & |pq| + t|zq| .
\end{eqnarray*}
By the law of cosines, we have
\begin{eqnarray*}
|zq|^2 & = & |pq|^2(2-\frac{1}{\lambda})
\end{eqnarray*}
which implies
\begin{eqnarray*}
d_G(p,q) & \leq & |pq| + t|pq|\sqrt{2-\frac{1}{\lambda}}.
\end{eqnarray*}
Therefore, we have to show that
\begin{eqnarray*}
t & \geq & \frac{1}{1-\sqrt{2-\frac{1}{\lambda}}}.
\end{eqnarray*}
Since $\beta\leq\theta$, we have $\cos\beta\geq\cos\theta$, and
thus
\begin{eqnarray*}
t & = & \frac{1}{(1-\lambda)\cos\theta}\\
  & \geq & \frac{1}{(1-\lambda)\cos\beta}\\
  & = & \frac{1}{(1-\lambda)(1/2\lambda)}\\
  & = & \frac{2\lambda}{1-\lambda}\\
  & \geq &\frac{1}{1-\sqrt{2-\frac{1}{\lambda}}} ,
\end{eqnarray*}
where the last inequality holds because it is equivalent to
$(1-\lambda)^2 \geq 0$.

\ \\
\textbf{Case 2.2: $\beta>\theta$.}  By the law of cosines we have
\begin{eqnarray*}
d_G(p,q) & \leq & |pq| + t|pq|\sqrt{2-2\cos\theta}
\end{eqnarray*}
which means that we have to show that
\begin{eqnarray*}
t & \geq & \frac{1}{1-\sqrt{2-2\cos\theta}} .
\end{eqnarray*}
But $\frac{1}{1-\sqrt{2-2\cos\theta}} \geq
\frac{1}{(1-\lambda)\cos\theta}$ since $\beta>\theta$ implies
$\cos\theta>\cos\beta=\frac{1}{2\lambda}$. This completes the last
case of the induction step. Note that the resulting $t$-spanning
paths found in this inductive proof are strong since both $|pr|$ and
$|rq|$ are shorter than $|pq|$. \end{proof}


The above proof provides a simple local routing algorithm. A routing
algorithm is considered local provided that the only information
used to make a decision is the 1-neighborhood of the current node as
well as the location of the destination (see~\cite{bose04a} for a
detailed description of the model). The routing algorithm proceeds
as follows. To find a path from $p$ to $q$, if the edge $(p,q)$ is
in $\paz(P)$, then take the edge. If the edge $(p,q)$ is not in
$\paz(P)$, then take an edge $(p,r)$ where $r$ is a destroyer of the
edge $(p,q)$. Recall that $r$ is a destroyer of the edge $(p,q)$ if
$r\in\overline{K}(p,q)$. This can be computed solely with the
positions of $p,q$ and $r$. Therefore, determining which of the
neighbors of $p$ in $\paz(P)$ destroyed the edge $(p,q)$ is a local
computation.


\subsection{$\paz$ is of Bounded Out-Degree}

\begin{figure}
\centering
\includegraphics{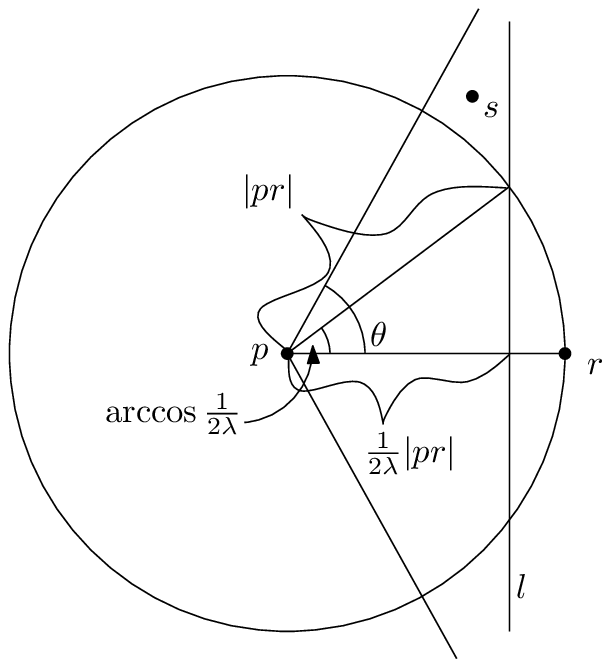}
\caption{The $\paz$ graph has bounded out-degree.}\label{fig-bounded}
\end{figure}

\begin{prop}The out-degree of a node in the $\paz$ graph is at most $\lfloor2\pi/\min(\theta, \arccos\frac{1}{2\lambda})\rfloor$.
\end{prop}
\begin{proof} Let $(p,r)$ and $(p,s)$ be two edges of the $\paz$ graph.
Without loss of generality, $|pr|\leq |ps|$. Let $l$ be the line
perpendicular to $\overline{pr}$ through
$p+\frac{1}{2\lambda}(r-p)$. Then either $\angle spr\geq \theta$ or
$s$ lies on the same side of $l$ as $p$. In the latter case, the
angle $\angle spr$ is at least $\arccos\frac{1}{2\lambda}$ (see
Figure~\ref{fig-bounded}). The angle $\angle spr$ is then at least
$\min(\theta, \arccos\frac{1}{2\lambda})$, which means that $p$ has
at most $\lfloor2\pi/\min(\theta, \arccos\frac{1}{2\lambda})\rfloor$
outgoing edges.\end{proof}

\vspace{-7mm}

\begin{cor} If $\theta \geq \pi/3$ and $\lambda > \frac{1}{2\cos(2\pi/7)}$,
then the out-degree of a node in the $\paz$ graph is at most six.
\end{cor}

\section{Half-Space Proximal}\label{section-comparison}

In this section, we outline the inconsistencies within statements of
the proof of the upper and lower bounds of HSP given in \citet{kranakis04}.

\begin{figure}[h]
\centering
\includegraphics{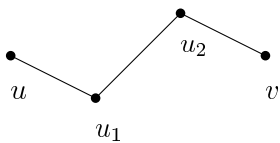}
\caption{Counter-example to the proof of Theorem 2 of
\cite{kranakis04}.}\label{fig-hsp-counter}
\end{figure}

In the proof of the upper bound (Theorem~2 of \citet{kranakis04}),
claim 4 states that \emph{all vertices $u_0, u_1, u_2, \ldots, u_k$ are either
in clockwise or anticlockwise order around $v$.}
The claim is that this situation
exists when $(u,v)$ is not an edge of HSP and no neighbor of $u$ is
adjacent to $v$.  However, as stated, this is not true. A
counter-example to this claim is shown in
Figure~\ref{fig-hsp-counter}.  There is a unique path from $u$ to
$v$, namely $uu_1u_2v$, but this path is neither clockwise nor
counter-clockwise around $v$. We believe that this situation may
exist {\em in the worst case}. However, a characterization of the
worst case situation must be given and it must be proven that the
worst case situation has the claimed property.

\begin{figure}[h]
\centering
\includegraphics[scale=0.9]{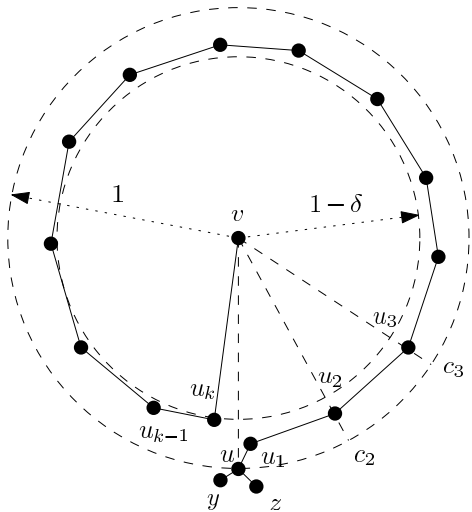}
\caption{The illustration of the lower bound on the spanning ratio
of~\cite{kranakis04}.}\label{fig-hsp-counter2}
\end{figure}

For the lower bound, the authors also claim that the stretch factor of HSP can
be arbitrarily close to $2\pi+1$. However, the proof they provide to support
that claim is a construction depicted in Figure~\ref{fig-hsp-counter2}
(reproduced from~\cite{kranakis04}). The claim is that the path from $u$ to $v$
can have length arbitrarily close to $(2\pi+1)|uv|$.  Although this may be true
for the path that they highlight. This path is {\em not} the only path from $u$
to $v$ in HSP.  The authors neglected the presence of the edge $(u_1u_k)$ in
their construction, which provides a shortcut that makes the distance between
$u$ and $v$ much less than~$2|uv|$.

One of the main reasons we believe the claims made in \citet{kranakis04} may be
true is that in the simulations, all the graphs have small stretch factor.
In fact, the stretch factor seems to be even smaller than $2\pi + 1$. However,
at this point, no proof that HSP is a constant spanner is known.
We provide a lower bound of $3-\epsilon$ on the
stretch factor of HSP as depicted in the construction below.

\begin{figure}[h]
\centering
\includegraphics{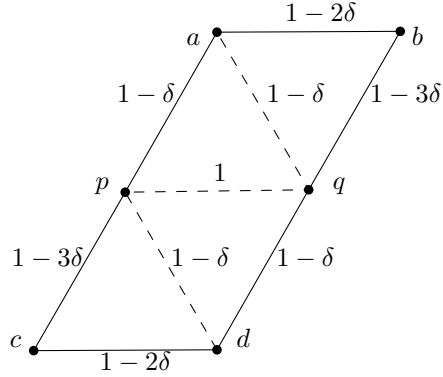}
\caption{Example of a 6 nodes HSP with a stretch factor of
$3-\epsilon$. The solid edges are in HSP.}\label{fig:lowerBound}
\end{figure}

\begin{prop}The HSP graph has stretch factor at least $3-\epsilon$.
\end{prop}
\begin{proof}
Consider the set of 6 point as in Figure~\ref{fig:lowerBound}, put
$\delta = \epsilon/6$. The length of the path between $p$ and $q$
via $a$ and $b$ is equal to the length of the path between $p$ and
$q$ via $c$ and $d$. The length of both of these paths is $3-
6\delta$. Since the shortest path between $p$ and $q$ in the HSP
graph is one of the above paths, the stretch factor is $3-6\delta= 3
- \epsilon$.
\end{proof}

\vspace{-5mm}

\section{Unit Disk Graph Spanners}\label{section-udg}

In Section~\ref{section-paz}, we showed that the $\paz$ graph of a
set of points in the plane is a strong $t$-spanner of the complete
graph of these points, for a constant $t=\frac{1}{(1- \lambda) \cos
\theta}$.  We show in this section that strong $t$-spanners are also
spanners of the unit disk graph. That is, the length of the shortest
path between a pair of points in the unit disk graph is not more
than $t$ times the length of the shortest path in the graph
resulting from the intersection of a strong $t$-spanner the a unit
disk graph. Before proceeding, we need to introduce some notation.

For simplicity of exposition, we will assume that given a set $P$ of
points in the plane, no two pairs of points are at equal distance
from each other. The \emph{complete geometric graph} defined on a
set $P$ of points, denoted $C(P)$, is the graph whose vertex set is
$P$ and whose edge set is $P\times P$, with each edge having its
weight equal to the Euclidean distance between its vertices. Let
$e_1,\ldots,e_{\binom{n}{2}}$ be the edges of $C(P)$ sorted
according to their lengths $L_1,\ldots,L_{\binom{n}{2}}$. For
$i=1\ldots\binom{n}{2}$, we denote by $C_i(P)$ the geometric graph
consisting of all edges whose length is no more than $L_i$. In
general, for any graph $G$ whose vertex set is $V$, we define $G_i$
as $G\cap C_i(V)$. Let $\UDG(P)$ be the unit disk graph of $P$,
which is the graph whose vertex set is $P$ and with edges between
pairs of vertices whose distance is not more than one. Note that
$\UDG(P)=C_i(P)$ for some~$i$.

We now show the relationship between strong $t$-spanners and unit
disk graphs.

\begin{prop}\label{prop1}
If $S$ is a strong $t$-spanner of $C(P)$, then for all
$i=1\ldots\binom{n}{2}$ and all $j=1\ldots i$, $S_i$ contains a
$t$-spanning path linking the vertices of $e_j$.
\end{prop}
\begin{proof}
Let $p$ and $q$ be the vertices of $e_j$.
Consider a strong $t$-spanner path in $S$ between $p$ and $q$.
Each edge on this path has length at most $|pq| = L_j \leq L_i$.
Therefore, each edge is in $S_i$.
\end{proof}
%
\begin{prop}\label{s-span}
If $S$ is a strong $t$-spanner of $C(P)$, then for all
$i=1\ldots\binom{n}{2}$, $S_i$ is a $t$-spanner of $C_i(P)$.
\end{prop}
\begin{proof} Let $a$ and $b$ be any two points such that
$d_{C_i(P)}(a,b)$ is finite. We need to show that in $S_i$ there
exists a path between $a$ and $b$
whose length is at most $t\cdot d_{C_i(P)}(a,b)$. Let
$a=p_1,p_2,\ldots,p_k=b$ be a shortest path in $C_i(P)$ between $a$
and $b$. Hence:
$$
d_{C_i(P)}(a,b) = \sum\limits_{j=1}^{k-1}|p_jp_{j+1}| .
$$
Now, by proposition~\ref{prop1}, for each edge $(p_j,p_{j+1})$ there
is a path in $S_i$ between $p_j$ and $p_{j+1}$ whose length is at
most $t\cdot |p_jp_{j+1}|$. Therefore:
$$
d_{S_i(P)}(a,b) \leq \sum\limits_{j=1}^{k-1}t\cdot |p_jp_{j+1}| =
t\sum\limits_{j=1}^{k-1}|p_jp_{j+1}| = t\cdot d_{C_i(P)}(a,b)
$$
which means that in $S_i$, there exists a path between $a$ and $b$
whose length is at most $t\cdot d_{C_i(P)}(a,b)$.
\end{proof}
%
\begin{cor}If $S$ is a strong $t$-spanner of $C(P)$, then $S \cap \UDG(P)$ is a strong $t$-spanner of $\UDG(P)$.
\end{cor}
\begin{proof} Just notice that $\UDG=C_i$ for some
$i$ and the result follows from Proposition \ref{s-span}.
\end{proof}

Thus, we have shown sufficient
conditions for a graph to be a spanner of the unit disk graph. We
now show that these conditions are also necessary.

\begin{prop}\label{prop-necessary}If $S$ is a subgraph of $C(P)$ such that for all $i=1\ldots\binom{n}{2}$,
$S_i$ is a $t$-spanner of
$C_i(P)$, then $S$ is a strong $t$-spanner of $C(P)$.
\end{prop}
\begin{proof} Let $a,b$ be any pair of points chosen in $P$. We have to show that in
$S$, there is a path between $a$ and $b$ such that
\begin{enumerate}
\item its length is at most $t\cdot |ab|$ and
\item every edge on the path has length at most $|ab|$.
\end{enumerate}

Let $e_i=(a,b)$. We know that $S_i$ is a $t$-spanner of $C_i(P)$.
Since $C_i(P)$ contains $e_i$, $d_{C_i(P)}(a,b)=|ab|$. Hence, there
is a path in $S_i$ (and therefore in $S$) whose length is at most
$t\cdot d_{C_i(P)}(a,b)=t|ab|$. Also, since it is in $S_i$, all of
its edges have length at most $L_i=|ab|$.
\end{proof}

\begin{figure}[h]
\centering
\includegraphics[scale=0.85]{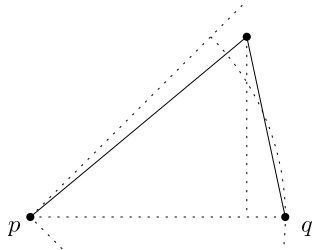}
\caption{The $\theta$-graph is not a strong
$t$-spanner.}\label{fig-theta-graph}
\end{figure}

The two last results, together, allow us to determine whether or not
given families of geometric graphs are also spanners of the unit
disk graph. First, since the $\paz$ graph is a strong $t$-spanner,
we already know that it is also a spanner of the unit disk graph.
Second, \citet{bose04} showed that the Yao graph\cite{Yao82} and the
Delaunay triangulation are strong $t$-spanners. Therefore, these
graphs are also spanners of the unit disk graph. Third, the
$\theta$-graph\cite{keil92} is not always a spanner of the unit disk
graph. The reason for that is that in a cone, the edge you chose may
not be the shortest edge. Hence, the path from a point $p$ to a
point $q$ may contain edges whose length is greater than $|pq|$ (see
Figure~\ref{fig-theta-graph}). Using
Proposition~\ref{prop-necessary}, we thus know that the intersection
of the $\theta$-graph with the unit disk graph may not be a spanner
of the unit disk graph. Indeed, the intersection of the
$\theta$-graph with the unit disk graph may not even be connected.



\section{Simulation Results}\label{section-simres}

In section~\ref{section-paz}, we provided worst-case analysis of the
spanning ratio of the $\paz$ graph. Using simulation, we now provide
estimates of the average spanning ratio of the $\paz$ graph. Using a
uniform distribution, we generated 200 sets of 200 points each and
computed the spanning ratio for $\lambda$ ranging from 0.5 to 1 and
$\theta$ ranging from $5^\circ$ to $90^\circ$ (for $\theta=0^\circ$,
the spanning ratio is exactly~1). For each graph, we then computed
the spanning ratio and the local routing ratio. The \emph{spanning
ratio} is defined as the maximum, over all pair of points $(p,q)$,
of the length of the shortest from $p$ to $q$ path in the $\paz$
graph divided by $|pq|$. The \emph{local routing ratio} is defined
as the maximum, over all pair of points $(p,q)$, of the length of
the path produced by using a local routing strategy in the $\paz$
graph divided by $|pq|$. The local routing strategy we have used is
the following: at each step, send the message to the neighbor which
destroyed $q$. We also tried the strategy which consists in choosing
the neighbor which is the nearest to $q$, and the results we
obtained were the same.

\begin{figure}
\begin{center}
\includegraphics[bb = 20 30 320 200]{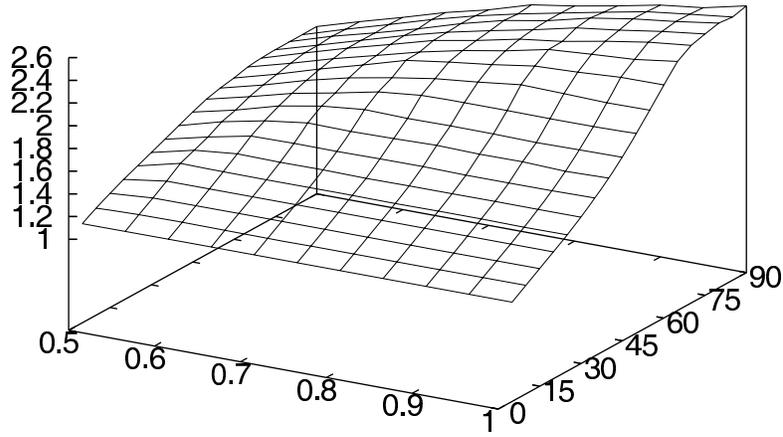}
\end{center}
\caption{Spanning Ratio for $\lambda=0.5$ to~1 and $\theta=5^\circ$
to $90^\circ$.}\label{fig-spanning}
\end{figure}

\begin{figure}
\begin{center}
\includegraphics[bb = 20 30 320 200]{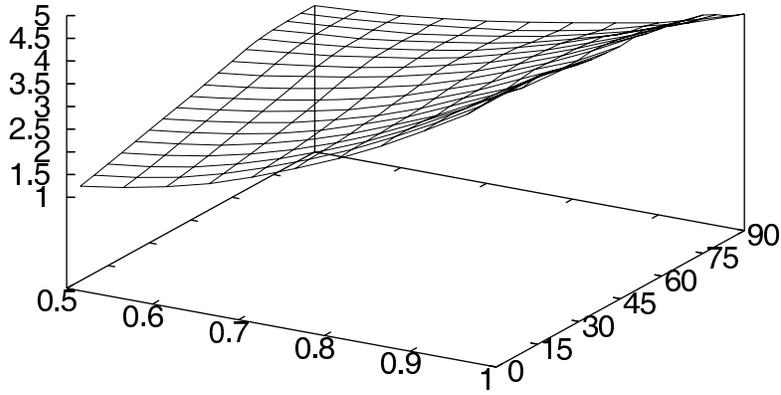}
\end{center}
\caption{Local Routing Ratio for $\lambda=0.5$ to~1 and
$\theta=5^\circ$ to $90^\circ$.}\label{fig-routing}
\end{figure}

Figure~\ref{fig-spanning} and Table~\ref{tab-spanning} show the
results we obtained for the spanning ratio. Figure~\ref{fig-routing}
and Table~\ref{tab-routing} show the results we obtained for the
local routing ratio. For the spanning ratio, the 95\% confidence
interval for these values is $\pm 0.0319$. For the local routing
ratio, the 95\% confidence interval is $\pm 0.0735$.

\begin{figure}
\begin{center}
\includegraphics[scale=0.8]{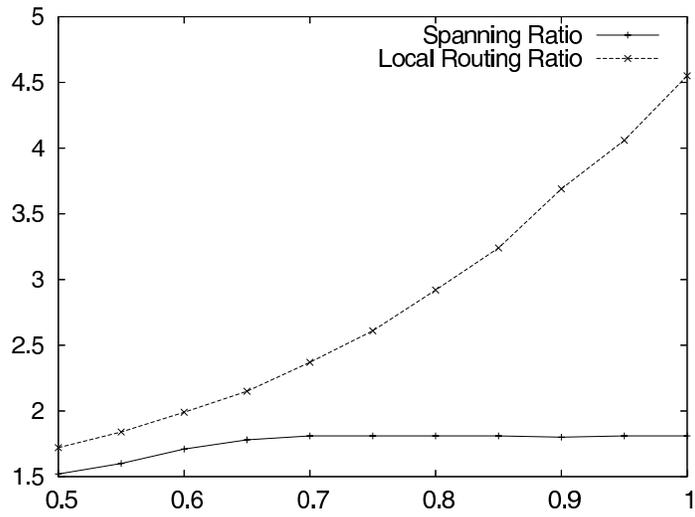}
\end{center}
\caption{Ratios for $\theta=45^\circ$.}\label{fig-theta}
\end{figure}

\begin{figure}
\begin{center}
\includegraphics[scale=0.8]{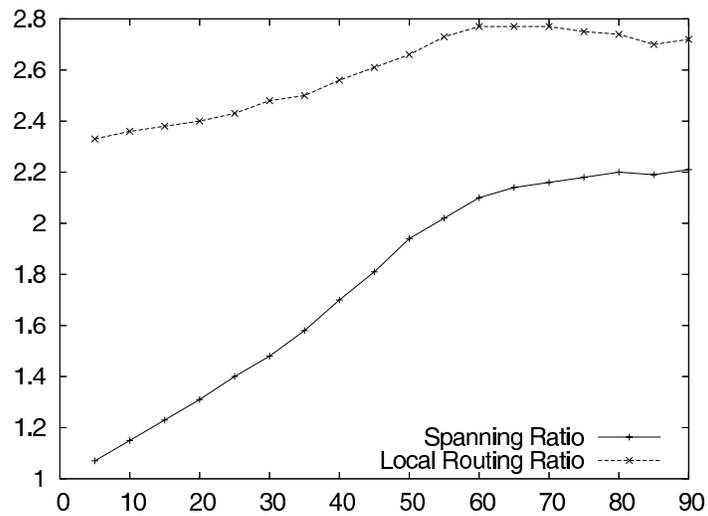}
\end{center}
\caption{Ratios for $\lambda=0.75$.}\label{fig-lambda}
\end{figure}

One interesting conclusion we can draw from these results is that
for the spanning ratio, $\theta$ has a more decisive influence than
$\lambda$. Figure~\ref{fig-theta} shows the simulation results for
the cases where $\lambda=0.75$. We see that even though both ratios
generally increase when $\theta$ increase, the spanning ratio varies
between 1.07 and 2.21 (107\% variation), while the local routing
ratio only varies between 2.33 and 2.77 (19\% variation).
For the local routing ratio, it is the other way around. It is
$\lambda$ which has a more decisive influence.
Figure~\ref{fig-lambda} shows the influence of $\lambda$ when
$\theta=45^\circ$. In that case, the local routing ratio varies
between 1.72 and 4.55 (165\% variation), while the spanning ratio
only varies between 1.52 and 1.81 (19\% variation).

\section{Conclusion}

We conclude with the problem that initiated this research. Determine
whether or not HSP is a strong $t$-spanner for some constant $t$.

\bibliographystyle{myBibliographyStyle}

\pagebreak

\begin{table}
\begin{tabular}{l|lllllllllll}
$\theta\setminus\lambda$ & 0.5 & 0.55 & 0.60 & 0.65 & 0.70 & 0.75 & 0.80 & 0.85 & 0.90 & 0.95 & 1\\
\hline
5  &  1.07 & 1.07 & 1.07 & 1.07 & 1.07 & 1.07 & 1.07 & 1.07 & 1.07 & 1.07 & 1.07\\
10 & 1.14 & 1.15 & 1.15 & 1.15 & 1.15 & 1.15 & 1.15 & 1.15 & 1.15 & 1.15 & 1.15\\
15 & 1.20 & 1.23 & 1.23 & 1.23 & 1.23 & 1.23 & 1.23 & 1.23 & 1.23 & 1.23 & 1.23\\
20 & 1.26 & 1.31 & 1.31 & 1.31 & 1.31 & 1.31 & 1.31 & 1.31 & 1.31 & 1.31 & 1.31\\
25 & 1.31 & 1.40 & 1.40 & 1.39 & 1.40 & 1.40 & 1.39 & 1.40 & 1.40 & 1.39 & 1.39\\
30 & 1.36 & 1.45 & 1.49 & 1.48 & 1.48 & 1.48 & 1.48 & 1.48 & 1.48 & 1.48 & 1.48\\
35 & 1.41 & 1.50 & 1.58 & 1.58 & 1.58 & 1.58 & 1.59 & 1.58 & 1.58 & 1.58 & 1.59\\
40 & 1.46 & 1.57 & 1.64 & 1.69 & 1.69 & 1.70 & 1.71 & 1.70 & 1.70 & 1.68 & 1.69\\
45 & 1.52 & 1.60 & 1.71 & 1.78 & 1.81 & 1.81 & 1.81 & 1.81 & 1.80 & 1.81 & 1.81\\
50 & 1.56 & 1.65 & 1.75 & 1.82 & 1.90 & 1.94 & 1.95 & 1.95 & 1.95 & 1.94 & 1.95\\
55 & 1.59 & 1.69 & 1.80 & 1.89 & 1.96 & 2.02 & 2.06 & 2.11 & 2.10 & 2.09 & 2.09\\
60 & 1.61 & 1.72 & 1.83 & 1.94 & 2.05 & 2.10 & 2.15 & 2.21 & 2.22 & 2.25 & 2.25\\
65 & 1.65 & 1.75 & 1.86 & 1.95 & 2.05 & 2.14 & 2.20 & 2.28 & 2.31 & 2.34 & 2.39\\
70 & 1.66 & 1.77 & 1.88 & 1.99 & 2.09 & 2.16 & 2.24 & 2.29 & 2.36 & 2.42 & 2.46\\
75 & 1.66 & 1.77 & 1.88 & 2.00 & 2.09 & 2.18 & 2.26 & 2.34 & 2.39 & 2.45 & 2.50\\
80 & 1.67 & 1.79 & 1.88 & 1.99 & 2.10 & 2.20 & 2.27 & 2.36 & 2.42 & 2.47 & 2.52\\
85 & 1.66 & 1.78 & 1.89 & 2.00 & 2.10 & 2.19 & 2.27 & 2.35 & 2.42 & 2.47 & 2.51\\
90 & 1.67 & 1.78 & 1.89 & 2.00 & 2.09 & 2.21 & 2.26 & 2.33 & 2.41 &
2.46 & 2.54
\end{tabular}
\caption{Spanning Ratio for $\lambda=0.5$ to~1 and $\theta=5^\circ$
to $90^\circ$.}\label{tab-spanning}
\end{table}

\begin{table}
\begin{tabular}{l|lllllllllll}
$\theta\setminus\lambda$ & 0.5 & 0.55 & 0.60 & 0.65 & 0.70 & 0.75 & 0.80 & 0.85 & 0.90 & 0.95 & 1\\
\hline
5 & 1.08 & 1.22 & 1.41 & 1.65 & 1.97 & 2.33 & 2.72 & 3.17 & 3.68 & 4.22 & 4.92\\
10 & 1.15 & 1.27 & 1.45 & 1.70 & 2.01 & 2.36 & 2.76 & 3.21 & 3.70 & 4.23 & 4.90\\
15 & 1.23 & 1.35 & 1.51 & 1.75 & 2.04 & 2.38 & 2.76 & 3.21 & 3.72 & 4.27 & 4.90\\
20 & 1.31 & 1.43 & 1.59 & 1.81 & 2.07 & 2.40 & 2.77 & 3.21 & 3.67 & 4.22 & 4.84\\
25 & 1.39 & 1.51 & 1.66 & 1.87 & 2.12 & 2.43 & 2.79 & 3.17 & 3.65 & 4.13 & 4.73\\
30 & 1.48 & 1.59 & 1.74 & 1.92 & 2.17 & 2.48 & 2.80 & 3.21 & 3.63 & 4.19 & 4.66\\
35 & 1.56 & 1.68 & 1.82 & 2.00 & 2.23 & 2.50 & 2.83 & 3.23 & 3.60 & 4.14 & 4.60\\
40 & 1.64 & 1.76 & 1.90 & 2.08 & 2.29 & 2.56 & 2.87 & 3.23 & 3.65 & 4.07 & 4.58\\
45 & 1.72 & 1.84 & 1.99 & 2.15 & 2.37 & 2.61 & 2.92 & 3.24 & 3.69 & 4.06 & 4.55\\
50 & 1.81 & 1.92 & 2.07 & 2.25 & 2.45 & 2.66 & 2.95 & 3.30 & 3.67 & 4.10 & 4.58\\
55 & 1.90 & 2.01 & 2.17 & 2.33 & 2.51 & 2.73 & 2.98 & 3.31 & 3.66 & 4.08 & 4.50\\
60 & 1.99 & 2.10 & 2.23 & 2.37 & 2.55 & 2.77 & 3.00 & 3.31 & 3.65 & 4.03 & 4.45\\
65 & 2.08 & 2.18 & 2.29 & 2.44 & 2.58 & 2.77 & 3.01 & 3.28 & 3.61 & 3.96 & 4.40\\
70 & 2.15 & 2.23 & 2.33 & 2.45 & 2.60 & 2.77 & 3.00 & 3.26 & 3.59 & 3.92 & 4.31\\
75 & 2.18 & 2.26 & 2.34 & 2.46 & 2.60 & 2.75 & 2.94 & 3.19 & 3.46 & 3.80 & 4.24\\
80 & 2.20 & 2.28 & 2.36 & 2.44 & 2.58 & 2.74 & 2.91 & 3.13 & 3.37 & 3.67 & 4.08\\
85 & 2.22 & 2.27 & 2.36 & 2.47 & 2.58 & 2.70 & 2.87 & 3.05 & 3.32 & 3.55 & 3.87\\
90 & 2.21 & 2.27 & 2.34 & 2.45 & 2.57 & 2.72 & 2.88 & 3.05 & 3.22 &
3.47 & 3.76
\end{tabular}
\caption{Local Routing Ratio for $\lambda=0.5$ to~1 and
$\theta=5^\circ$ to $90^\circ$.}\label{tab-routing}
\end{table}

\end{document}